\documentclass[12pt]{article}
\usepackage{amsmath}
\usepackage{cite}
\usepackage{a41}

\newcommand\Mvec{\,\mbox{\bf M}}
\def\z#1{{\zeta_{#1}}}
\def\Li#1{{\rm Li}_{#1}}
\def\ttD#1{{\tt D}$--$#1}
\def\S(#1){S_{#1}(N)}
\def\lnt{\ln(2)}

\begin{document}
\sloppy
\setlength{\baselineskip}{0.515cm}
\thispagestyle{empty}
\begin{flushleft}
DESY 05--007 \hfill
{\tt hep-ph/0503188}\\
SFB/CPP--05--05\\
March 2005\\
\end{flushleft}

\setcounter{page}{0}

\mbox{}
\vspace*{\fill}
\begin{center}

{\LARGE\bf Analytic Continuation of the Harmonic Sums}  \\
\vspace*{3mm}

{\LARGE\bf for the 3--Loop Anomalous Dimensions}\\

\vspace{4cm}
\large
Johannes Bl\"umlein and Sven-Olaf Moch

\vspace{1.5cm}
\normalsize
{\it  Deutsches Elektronen--Synchrotron, DESY,}\\
{\it  Platanenallee 6, D-15735 Zeuthen, Germany}
\\
%%\today

\end{center}
\normalsize
\vspace{\fill}
\begin{abstract}
\noindent
We present for numerical use the analytic continuations to complex arguments 
of those basic Mellin transforms, which build the harmonic sums contributing 
to the 3--loop anomalous dimensions. Eight new basic functions contribute 
in addition to the analytic continuations for the 2--loop massless Wilson 
coefficients calculated previously. The representations derived have a 
relative accuracy of better than $10^{-7}$ in the range 
$x~\epsilon~[10^{-6},0.98]$.
\end{abstract}

\vspace{1mm}
\noindent

\vspace*{\fill}
\noindent
\numberwithin{equation}{section}
%%%%%%%%%%%%%%%%%%%%%%%%%%%%%%%%%%%%%%%%%%%%%%%%%%%%%%%%%%%%%%%%%%%%%%%
%%%%%%%%%%%%%%%%%%%%%%%%%%%%%%%%%%%%%%%%%%%%%%%%%%%%%%%%%%%%%%%%%%%%%%%%
\newpage
\section{Introduction}
%%%%%%%%%%%%%%%%%%%%%%%%%%%%%%%%%%%%%%%%%%%%%%%%%%%%%%%%%%%%%%%%%%%%%%%%
%

\vspace{1mm}\noindent
A very efficient way to calculate the evolution of the parton densities 
in deeply inelastic scattering consist in solving the evolution equations
in Mellin--$N$ space~\cite{Gluck:1989ze,Blumlein:1998em}. 
With the advent of the 3--loop anomalous 
dimensions~\cite{Moch:2004pa,Vogt:2004mw}
the scaling violations of the deep--inelastic structure 
functions can be analyzed at high precision. 
The evolution equations are ordinary differential equations,
which can be solved analytically. 
The Wilson coefficients~\cite{
Kazakov:1987jk,SanchezGuillen:1990iq,vanNeerven:1991nn,%Zijlstra:1991qc,%
%Zijlstra:1992qd,Zijlstra:1992kj,
Moch:1999eb}
and anomalous dimensions are given in terms of multiple nested harmonic 
sums~\cite{Vermaseren:1998uu,Blumlein:1998if},
which obey algebraic relations~\cite{Vermaseren:1998uu,Blumlein:1998if,Borwein:1996,Blumlein:2003gb}.
Each finite harmonic sum can be represented in terms of a Mellin transform of a 
harmonic polylogarithm~\cite{Remiddi:1999ew} weighted by $1/(1 \pm x)$ and a polynomial of harmonic 
sums of lower weight. 
It turns out that for all basic functions at least up to weight five,
as needed for the physical observables mentioned above,
the harmonic polylogarithms are Nielsen integrals \cite{Nielsen,Kolbig:1983qt}.
Furthermore, structural relations exist between various Mellin transforms of 
weighted Nielsen integrals~\cite{Blumlein:2004bb}. 
They are due to relations at rational values of the Mellin variable, 
differentiation w.r.t. the Mellin variable, integration by parts relations 
and functional identities of Nielsen integrals. 
These relations, together with the quasi-shuffle algebra of harmonic sums 
allow to build the massless single--scale quantities in perturbative QED and 
QCD out of a low number of basic functions at a given order in
perturbation theory.

The anomalous dimensions and Wilson coefficients in Mellin space, for instance, 
can be written in terms of these basic functions and a number of their derivatives, 
in addition to polynomials including Euler's $\psi$--functions as well as rational
functions of the Mellin variable for complex values of $N$.
Then, on this basis, results in $x$--space, are obtained by means of a single inverse 
Mellin transform, which has to be carried out numerically.
For the class of functions contributing to the 2--loop Wilson coefficients 
the analytic continuations were given in Ref.~\cite{Blumlein:2000hw}.
Simple analytic continuations of nested sums are known for long, cf. e.g. Ref.~\cite{Blumlein:1997vf}. 
This direction has been followed in Ref.~\cite{Kotikov:2005gr} recently. However, in many cases the 
numerical convergence is very slow. 

To obtain an appropriate representation for the 3--loop anomalous dimensions
and 2--loop Wilson coefficients,
we exploit the algebraic and structural relations between the nested harmonic sums.
For the 2--loop anomalous dimensions only one non--trivial function contributes 
aside from Euler's $\psi^{(k)}(N)$ 
functions~\cite{Gonzalez-Arroyo:1979df,Gonzalez-Arroyo:1980he}. 
All known massless 2--loop Wilson 
coefficients can be expressed by four more
functions~\cite{JBSM1,Blumlein:2005im}. 
For the 3--loop anomalous dimension the Mellin transforms 
%--------------------------------------------------------------------------
\begin{equation}
\Mvec[f(x)](N) = \int_0^1~dx~x^N~f(x)
\end{equation}
%--------------------------------------------------------------------------
of the following eight functions form the set of additional basic 
functions~: 
%--------------------------------------------------------------------------
\begin{eqnarray}
{\Li4(x) \over 1 \pm  x},
\hspace*{7mm}
{S_{1,3}(x) \over 1+x},
\hspace*{7mm}
{S_{2,2}(x) \over 1 \pm x},
\hspace*{7mm}
{\Li2^2(x) \over 1+x},
\hspace*{7mm}
{S_{2,2}(-x) - \Li2^2(-x)/2 \over 1 \pm x}~.
\end{eqnarray}
%------------------------------------------------------------------------------------------------
Here the functions ${\rm Li}_n(x)$ and $S_{p,n}(x)$ denote 
polylogarithms~\cite{lewin:book} and Nielsen--integrals \cite{Nielsen,Kolbig:1983qt},
respectively.
In terms of harmonic polylogarithms~\cite{Remiddi:1999ew}, they read
%\begin{eqnarray}
\begin{alignat}{4}
\Li4(x) &= H_{0,0,0,1}(x), \hspace{5mm} & %\qquad
S_{1,3}(x) &= H_{0,1,1,1}(x), \hspace{5mm} & %\qquad
S_{2,2}(x) &= H_{0,0,1,1}(x), & %\qquad
\nonumber \\
\Li2^2(x) &= (H_{0,1}(x))^2,  \hspace{5mm} &
S_{2,2}(-x) &= H_{0,0,-1,-1}(x),  \hspace{5mm} &
\Li2^2(-x) &= (-H_{0,-1}(x))^2, &
%\end{eqnarray}
\end{alignat}
where the algebra can be applied to products of harmonic polylogarithms to express 
them in the basis of single harmonic polylogarithms.

In the present paper we derive fast and precise numerical 
representations of these Mellin transforms for complex values of $N$.
Previously numerical representations of the 3--loop anomalous dimensions were given
in \cite{Moch:2004pa,
Vogt:2004mw}.~\footnote{Related 
earlier approximate representations were given in 
\cite{vanNeerven:1999ca}%%,%
%vanNeerven:2000uj,vanNeerven:2000wp,vanNeerven:2001pe}
%}
.} 
Following earlier investigations~\cite{Blumlein:2000hw}, we aim at numerical
representations for the individual basic functions, the building 
blocks of the massless Wilson coefficients and the anomalous dimensions, 
to high precision.
Thus, in principle, the present approach is tunable to arbitrary accuracy.
%%%%%%%%%%%%%%%%%%%%%%%%%%%%%%%%%%%%%%%%%%%%%%%%%%%%%%%%%%%%%%%%%%%%%%%%
%\newpage
\section{Mellin Transforms of the Type \boldmath{$f(x)/(1+x)$}}
%%%%%%%%%%%%%%%%%%%%%%%%%%%%%%%%%%%%%%%%%%%%%%%%%%%%%%%%%%%%%%%%%%%%%%%%

\vspace{1mm}\noindent
As outlined in Ref.~\cite{Blumlein:2000hw} the Mellin transforms containing the 
denominator $1/(1+x)$ are given by:
%--------------------------------------------------------------------------
\begin{eqnarray}
\label{eq11}
\Mvec\left[\frac{f(x)}{1+x}\right](N) &=& \lnt \cdot f(1)
- \int_0^1 dx ~x^{N-1}~\ln(1+x) \left[N f(x) + x f'(x)\right]
\nonumber\\ &=& \lnt \cdot f(1) - \sum_{k=1}^{20} a^{(1)}_k \left(
N \Mvec[f(x)](N+k-1) + \Mvec\left[f'(x)\right](N+k)\right)
\, .
\nonumber\\
\end{eqnarray}
%--------------------------------------------------------------------------
Accurate representations for the function $\ln(1+x)$ can be given using the  
minimax method for numerical approximation:
%--------------------------------------------------------------------------
\begin{eqnarray}
\ln(1+x) \simeq \sum_{k=1}^{20} a^{(1)}_k x^k
\end{eqnarray}
%--------------------------------------------------------------------------
which we further improved to an accuracy of $3 \times 10^{-16}$. The 
coefficients 
read:
%-----------------------------------------------------------------------
\small
\begin{eqnarray}
%NEW
\begin{array}{lcrlcr}
%%STARTMAPLE
%% - \ln(1+x) +
%%{0
a^{(1)}_{1}   &\:=\:&    0.9999999999999925\ttD0 &
a^{(1)}_{2}   &\:=\:&   -0.4999999999988568\ttD0
\nonumber\\
a^{(1)}_{3}   &\:=\:&    0.3333333332641123\ttD0 &
a^{(1)}_{4}   &\:=\:&   -0.2499999977763199\ttD0
\nonumber\\
a^{(1)}_{5}   &\:=\:&    0.1999999561535526\ttD0 &
a^{(1)}_{6}   &\:=\:&   -0.1666660875051348\ttD0 
\nonumber\\
a^{(1)}_{7}   &\:=\:&    0.1428517138099479\ttD0 &
a^{(1)}_{8}   &\:=\:&   -0.1249623936313475\ttD0  
\nonumber\\
a^{(1)}_{9}   &\:=\:&    0.1109128496887138\ttD0 &
a^{(1)}_{10}   &\:=\:&   -0.9918652787800788\ttD1 
\nonumber\\
a^{(1)}_{11}   &\:=\:&   0.8826572954250856\ttD1 &
a^{(1)}_{12}   &\:=\:&  -0.7643209265133132\ttD1 
\nonumber\\
a^{(1)}_{13}   &\:=\:&   0.6225829212455825\ttD1 &
a^{(1)}_{14}   &\:=\:&  -0.4572477090315515\ttD1
\nonumber\\
a^{(1)}_{15}   &\:=\:&   0.2890194939889559\ttD1 &
a^{(1)}_{16}   &\:=\:&  -0.1496621145891488\ttD1
\nonumber\\
a^{(1)}_{17}   &\:=\:&   0.6003156359511387\ttD2 &
a^{(1)}_{18}   &\:=\:&  -0.1731328252868496\ttD2
\nonumber\\
a^{(1)}_{19}   &\:=\:&   0.3172112728405899\ttD3 &
a^{(1)}_{20}   &\:=\:&  -0.2760099875146713\ttD4
%%};
%%STOPMAPLE
\end{array}
\nonumber
\end{eqnarray}
\normalsize
%-----------------------------------------------------------------------
For later use we also parameterize $\ln^2(1+x)$
%--------------------------------------------------------------------------
\begin{eqnarray}
\ln^2(1+x) \simeq \sum_{k=2}^{24} a^{(2)}_k x^k~,
\end{eqnarray}
%--------------------------------------------------------------------------

with
%--------------------------------------------------------------------------
\small
\begin{eqnarray}
%NEW
\begin{array}{lcrlcr}
%%STARTMAPLE
%% - \ln(1+x)^2 + 
%%{0
a^{(2)}_{2}   &\:=\:&   1.0000000000000000\ttD0 &
a^{(2)}_{3}   &\:=\:&  -0.9999999999999985\ttD0   
\nonumber\\
a^{(2)}_{4}   &\:=\:&   0.9166666666663948\ttD0 &
a^{(2)}_{5}   &\:=\:&  -0.8333333333136118\ttD0
\nonumber\\
a^{(2)}_{6}   &\:=\:&   0.7611111103508889\ttD0 &
a^{(2)}_{7}   &\:=\:&  -0.6999999819735105\ttD0
\nonumber\\
a^{(2)}_{8}   &\:=\:&   0.6482139985629993\ttD0 &
a^{(2)}_{9}   &\:=\:&  -0.6039649964806160\ttD0
\nonumber\\
a^{(2)}_{10}   &\:=\:&    0.5657662306410356\ttD0 &
a^{(2)}_{11}   &\:=\:&   -0.5323631718571445\ttD0
\nonumber\\
a^{(2)}_{12}   &\:=\:&   0.5024238774786239\ttD0 &
a^{(2)}_{13}   &\:=\:&  -0.4738508288315496\ttD0
\nonumber\\
a^{(2)}_{14}   &\:=\:&   0.4427472719775835\ttD0 &
a^{(2)}_{15}   &\:=\:&  -0.4029142806330511\ttD0
\nonumber\\
a^{(2)}_{16}   &\:=\:&   0.3476841543351489\ttD0 &
a^{(2)}_{17}   &\:=\:&  -0.2748590021353420\ttD0
\nonumber\\
a^{(2)}_{18}   &\:=\:&   0.1915627642585285\ttD0 &
a^{(2)}_{19}   &\:=\:&  -0.1130763066428224\ttD0
\nonumber\\
a^{(2)}_{20}   &\:=\:&   0.5415661067306229\ttD1 &
a^{(2)}_{21}   &\:=\:&  -0.1999877298940919\ttD1
\nonumber\\
a^{(2)}_{22}   &\:=\:&   0.5303624439388411\ttD2 &
a^{(2)}_{23}   &\:=\:&  -0.8944156375768203\ttD3
\nonumber\\
a^{(2)}_{24}   &\:=\:&    0.7179502917974332\ttD4 &
               & &  
%%};
%%STOPMAPLE
\end{array}
\nonumber
\end{eqnarray}
\normalsize

%-----------------------------------------------------------------------
The following Mellin transforms contribute, cf. 
also Refs.~\cite{Vermaseren:1998uu,Blumlein:1998if,Blumlein:2000hw}, 
%-----------------------------------------------------------------------
\begin{eqnarray}
\Mvec\left[\Li4(x)\right](N-1) &=& 
%%STARTFORM
%%L %%MLi4x =
	  {\z4 \over N} - {\z3 \over N^2} + {\z2 \over N^3} - {\S(1) \over N^4}
%%;
%%STOPFORM
\\
\Mvec\left[\Li4'(x)\right](N-1) &=& \Mvec\left[\Li3(x)\right](N-2)
\\
\Mvec\left[S_{1,3}(x)\right](N-1) &=& 
%%STARTFORM
%%L %%MS13x =
	  {\z4 \over N} - {(\S(1))^3 \over 6 \* N^2} 
	  - {\S(1) \* \S(2) \over 2 \* N^2} 
	  - {\S(3) \over 3 \* N^2} 
%%;
%%STOPFORM
\\
\Mvec\left[S_{1,3}'(x)\right](N-1) &=& 
          - {1 \over 6} \* \Mvec\left[\ln^3(1-x)\right](N-2)
\\
\Mvec\left[\Li2^2(x)\right](N-1) &=& 
%%STARTFORM
%%L %%MLi2x2 =
	  {1 \over N} \* \left( \z2^2 - {4 \* \z3 \over N} 
	  - 2 \* \z2 \* {\S(1) \over N} 
	  + 2 \* {\S(2,1) \over N} \right) 
\nonumber\\
& &
          + {2 \over N^3} \* \left( (\S(1))^2 + \S(2) \right)
%%;
%%STOPFORM
\\
\Mvec\left[(\Li2^2(x))'\right](N-1) &=& 
          - 2 \* \Mvec\left[\ln(1-x) \Li2(x)\right](N-2) 
\\
\Mvec\left[S_{2,2}(x)\right](N-1) &=& 
%%STARTFORM
%%L %%MS22x =
	  {\z4 \over 4 \* N} - {\z3 \over N^2} + {(\S(1))^2 + \S(2) \over 2 \* N^3}
%%;
%%STOPFORM
\\
\Mvec\left[S_{2,2}'(x)\right](N-1) &=&  \Mvec\left[S_{1,2}(x)\right](N-2) 
\\
\Mvec\left[S_{2,2}(-x)-\Li2^2(-x)/2\right](N-1) &=& 
%%STARTFORM
%%L %%MS22mxmLi2mx2o2 =
	  {1 \over N} \* \biggl[ - {3 \over 4} \* \z2^2 + 2 \* \Li4(1/2) 
	  + {7 \over 4} \* \z3 \* \lnt  - {1 \over 2} \* \z2 \* (\lnt)^2 
\nonumber\\
& &
          + {1 \over 12} \* (\lnt)^4 - 
%%STOPFORM
\Mvec\left[S_{1,2}(-x)\right](N-1)
%%STARTFORM
%%MMS12mx
          \biggr]
\nonumber\\ & &
          - {1 \over N} \* \left[ {\z2^2 \over 8} + 
%%STOPFORM
\Mvec\left[\ln(1+x)\Li2(-x)\right](N-1) 
%%STARTFORM
%%MMln1pxLimx
          \right]
%%;
%%STOPFORM
\nonumber\\
\\
\Mvec\left[\left(S_{2,2}(-x)-\Li2^2(-x)/2\right)'\right](N-1) &=&  
          \Mvec\left[S_{1,2}(-x)\right](N-2) 
\nonumber\\
& &
          +  \Mvec\left[\ln(1+x) \Li2(-x)\right](N-2)
\end{eqnarray}
%-----------------------------------------------------------------------
with
%-----------------------------------------------------------------------
\begin{eqnarray}
\Mvec\left[\Li3(x)\right](N-1) &=&
%%STARTFORM
%%L %%MLi3x = 
	  {\z3 \over N} - {\z2 \over N^2} + {\S(1) \over N^3}
%%;
%%STOPFORM
\\
\Mvec\left[\ln^3(1-x)\right](N-1) &=& 
%%STARTFORM
%%L %%Mln31mx =
          - {(\S(1))^3 \over N} 
	  - 3 \* {\S(1) \* \S(2) \over N} 
	  - 2 \* {\S(3) \over N}
%%;
%%STOPFORM
\\
 \Mvec\left[S_{1,2}(x)\right](N-1) &=& 
%%STARTFORM
%%L %%MS12x =
	   {\z3 \over N} 
	   - {(\S(1))^2 \over 2 \* N^2} 
	   - {\S(2) \over 2 \* N^2}
%%;
%%STOPFORM
\\
\Mvec\left[S_{1,2}(-x)\right](N-1) &=& 
%%STARTFORM
%%L %%MS12mx =
	  {\z3 \over 8 \* N} - {(\lnt)^2 \over 2 \* N^2}
	  - {(-1)^N \over N^2} \* \biggl[ \S(1,-1) 
	  + \lnt \* \left( \S(1) - \S(-1) \right) 
\nonumber\\
& & 
          - {1 \over 2} \* (\lnt)^2 \biggr]
%%;
%%STOPFORM
\nonumber\\ 
&\simeq& 
          {\z3 \over 8 \* N} 
	  - {1 \over 2 \* N} \sum_{k=2}^{24} {a^{(2)}_k \over N+k}
\\
\Mvec\left[\ln(1-x) \Li2(x)\right](N-1) &=& 
%%STARTFORM
%%L %%Mln1mxLix = 
	  {1 \over N} \* \biggl[ - 2 \* \z3 - \z2 \* \S(1) 
	  + {1 \over N} \* \left((\S(1))^2 + \S(2) \right) 
	  + \S(2,1) \biggr]
\nonumber\\ 
%%;
%%STOPFORM
\end{eqnarray}\begin{eqnarray}
\Mvec\left[\ln(1+x) \Li2(-x)\right](N-1) &\simeq& 
          \sum_{k=1}^{20} a^{(1)}_k \Mvec\left[\Li2(-x)\right](N-1+k)
\\
\Mvec\left[\Li2(-x)\right](N-1) &=& 
%%STARTFORM
%%L %%MLi2mx =
          - {\z2 \over 2 \* N} + {\lnt \over N^2} 
	  - {(-1)^N \over N^2} \* (\S(-1) + \lnt )
%%;
%%STOPFORM
%\nonumber\\
\end{eqnarray}
%-----------------------------------------------------------------------
The functions $\S(1,-1)$ and $\S(2,1)$ have been parameterized in Ref.~\cite{Blumlein:2000hw}
%-----------------------------------------------------------------------
\begin{eqnarray}
\S(1,-1) &=& (-1)^{N+1} \Mvec\left[\frac{\ln(1+x)}{1+x}\right](N) -
\lnt \left(\S(1) - \S(-1) \right) + \frac{1}{2} (\lnt)^2\\
\S(2,1) &=& \Mvec\left[\left(\frac{\Li2(x)}{1-x}\right)_+\right](N) + \z2 \* \S(1)
\, ,
\end{eqnarray}
%------------------------------------------------------------------------
where 
%-----------------------------------------------------------------------
\begin{eqnarray}
\Mvec\left[\frac{\ln(1+x)}{1+x}\right](N) \simeq 
\frac{1}{2} (\lnt)^2 - \frac{N}{2}\sum_{k=2}^{24} 
\frac{a^{(2)}_k}{N+k}~.
\end{eqnarray}
%------------------------------------------------------------------------
%%%%%%%%%%%%%%%%%%%%%%%%%%%%%%%%%%%%%%%%%%%%%%%%%%%%%%%%%%%%%%%%%%%%%%%%
%\newpage
\section{Mellin Transforms of the Type 
\boldmath{$\left(f(x)/(1-x)\right)_+$}}
%%%%%%%%%%%%%%%%%%%%%%%%%%%%%%%%%%%%%%%%%%%%%%%%%%%%%%%%%%%%%%%%%%%%%%%%

\vspace{1mm} \noindent
Three functions of this type contribute. In case the numerator function 
is analytic at $x=1$ one may represent the functions of this class using
the minimax method directly. Both $\Li4(x)$ and $S_{2,2}(x)$ have 
branch points at $x=1$. Therefore we derive an analytic representation 
$\hat{f}_i(x)$ for $\left[f_i(x)-f_i(1)\right]/(1-x)$ valid for the region 
$x \rightarrow 1$ and determine the remainder part using the 
minimax method:
%-----------------------------------------------------------------------------
\begin{eqnarray}
\label{eqR1}
\frac{f_i(x) -f_i(1)}{1-x} = 
\hat{f}_i(x) +
\sum_{k=0}^{k_{\rm max}} c_k^{(i)} x^k~. 
\end{eqnarray}
%------------------------------------------------------------------------------

For $[\Li4(x) - \z4]/(1-x)$ one obtains
%-----------------------------------------------------------------------
\begin{eqnarray}
\hat{f}_1(x) &=& 
%%STARTMAPLE
%%L %%f1func = 
- \z3 + {1 \over 2} \* (\z2-\z3) \* (1-x) + \left({1 \over 6} \* \ln(1-x) 
+ {\z2 \over 2} - {11 \over 36} - {\z3 \over 3} \right) \* (1-x)^{2}
\nonumber\\ &&
+ \left( {1 \over 4} \* \ln(1-x) +  {11 \over 24} \* \z2 
- {19 \over 48} - {\z3 \over 4} \right) \* (1-x)^{3}
\nonumber\\ &&
+ \left( {7 \over 24} \* \ln(1-x) + {5 \over 12} \* \z2 - {\z3 \over 5} 
 - {599 \over 1440} \right) \* (1-x)^{4}
\nonumber\\ &&
+ \left( {5 \over 16} \* \ln(1-x) + {137 \over 360} \* \z2 
- {79 \over 192} -{\z3 \over 6} \right) \* (1-x)^{5}
%%;
%%STOPMAPLE
\end{eqnarray}

\small
\begin{eqnarray}
%NEW
\begin{array}{lcrlcr}
%%STARTMAPLE
%%L %%f1appr = 
%%{0
c^{(1)}_{0}   &\:=\:&  -3.187045493829754\ttD1 &
c^{(1)}_{1}   &\:=\:&   1.752102582962004\ttD0
\nonumber\\
c^{(1)}_{2}   &\:=\:&  -3.926780960319761\ttD0 &
c^{(1)}_{3}   &\:=\:&   4.533622455411171\ttD0
\nonumber\\
c^{(1)}_{4}   &\:=\:&  -2.764070067739643\ttD0 &
c^{(1)}_{5}   &\:=\:&   7.705635337822153\ttD1
\nonumber\\
c^{(1)}_{6}   &\:=\:&  -1.412597571664758\ttD2 &
c^{(1)}_{7}   &\:=\:&  -1.024735147728843\ttD1 
\nonumber\\
c^{(1)}_{8}   &\:=\:&   2.072912823276118\ttD1 &
c^{(1)}_{9}   &\:=\:&  -3.094485142894180\ttD1
\nonumber\\
c^{(1)}_{10}   &\:=\:&  3.100508803799690\ttD1 &
c^{(1)}_{11}   &\:=\:& -2.017510797419543\ttD1
\nonumber\\
c^{(1)}_{12}   &\:=\:&   7.682650942255444\ttD2 &
c^{(1)}_{13}   &\:=\:&  -1.310258217741916\ttD2
%%};
%%STOPMAPLE
\end{array}
\nonumber
\end{eqnarray}
\normalsize

\noindent
with $k_{\rm max} = 13.$ The Mellin transform of $\hat{f}_i(x)$ is given 
by a linear combination of 
%-----------------------------------------------------------------------------
\begin{eqnarray}
\Mvec\left[(1-x)^M\right](N) &=& B(N+1,M+1) \\
\Mvec\left[(1-x)^M \ln(1-x) \right](N) &=& \left[\psi(M+1) - 
\psi(M+N+2)\right] B(N+1,M+1) \\
\Mvec\left[(1-x)^M \ln^2(1-x) \right](N) &=& 
\bigl\{\left[\psi'(M+1) - \psi'(M+N+2)\right] 
\nonumber\\ &+& 
\left[\psi(M+1) - \psi(M+N+2)\right]^2\bigr\} B(N+1,M+1)~. 
\end{eqnarray}
%-----------------------------------------------------------------------------

$[S_{2,2}(x)-\z4/4)]/(1-x)$ obeys the 
representation (\ref{eqR1}), where 
%-----------------------------------------------------------------------
\begin{eqnarray}
\hat{f}_2(x) &=& 
%%STARTMAPLE
%%L %%f2func = 
(1-x) \* \biggl[ \left( 
- {7 \over 40} \* x^{5} + {767 \over 720} \* x^{4} - {979 \over 360} \* x^{3} 
+ {899 \over 240} \* x^{2} - {1069 \over 360} \* x + {469 \over 360} \right) \* \ln(1-x)^2 
\nonumber \\
& & +
\left(
{947 \over 3600} \* x^{5} - {11683 \over 7200} \* x^{4} 
+ {15221 \over 3600} \* x^{3} - {4827 \over 800} \* x^{2} 
+ {18511 \over 3600} \* x - {409 \over 150} \right) \* \ln(1-x)
\nonumber\\ & & 
- {104641 \over 504000} \* x^5 + {11675141 \over 9072000} \* x^4
- {15334867 \over 4536000} \* x^3 + {14820287 \over 3024000} \* x^2
\nonumber\\ & &
- {19680697 \over 4536000} \* x + {2964583 \over 1134000}
\biggr]
\nonumber\\ & & +
\left( 
- {1 \over 7} \* x^{6} + {43 \over 42} \* x^{5} - {667 \over 210} \* x^{4} + {2341 \over 420} \* x^{3} 
-  {853 \over 140} \* x^{2} +  {617 \over 140} \* x -  {363 \over 140} \right) \* \z3 
%%;
%%STOPMAPLE
\end{eqnarray}
%-----------------------------------------------------------------------
%-----------------------------------------------------------------------
\small
\begin{eqnarray}
%NEW
\begin{array}{lcrlcr}
%%STARTMAPLE
%%L %%f2appr = 
%%{0
c^{(2)}_{0}   &\:=\:&   2.319102959447070\ttD1 &
c^{(2)}_{1}   &\:=\:&  -1.341835247346528\ttD0
\nonumber\\
c^{(2)}_{2}   &\:=\:&   3.141213505948200\ttD0 &
c^{(2)}_{3}   &\:=\:&  -3.689298402805891\ttD0
\nonumber\\
c^{(2)}_{4}   &\:=\:&   2.066708069184852\ttD0 &
c^{(2)}_{5}   &\:=\:&  -0.221726622455410\ttD0
\nonumber\\
c^{(2)}_{6}   &\:=\:&  -2.866874606436715\ttD1 &
c^{(2)}_{7}   &\:=\:&   1.886633904893633\ttD1
\nonumber\\
c^{(2)}_{8}   &\:=\:&  -5.104536639955758\ttD1 &
c^{(2)}_{9}   &\:=\:&   1.814194435712243\ttD0
\nonumber\\
c^{(2)}_{10}   &\:=\:&  -4.810364434281610\ttD0 &
c^{(2)}_{11}   &\:=\:&   9.670512224938253\ttD0
\nonumber\\
c^{(2)}_{12}   &\:=\:&  -14.74512233105934\ttD0 &
c^{(2)}_{13}   &\:=\:&   16.93126634317140\ttD0
\nonumber\\
c^{(2)}_{14}   &\:=\:&  -14.39776586765419\ttD0 &
c^{(2)}_{15}   &\:=\:&    8.789666564310812\ttD0
\nonumber\\
c^{(2)}_{16}   &\:=\:&  -3.642212170527524\ttD0 &
c^{(2)}_{17}   &\:=\:&   9.173661876074224\ttD1
\nonumber\\
c^{(2)}_{18}   &\:=\:&  -1.060348165292345\ttD1 &
  & & 
%%};
%%STOPMAPLE
\end{array}
\nonumber
\end{eqnarray}
\normalsize
and $k_{\rm max} = 18$. 

The Nielsen integrals $S_{n,p}(x)$ are analytic at $x = -1$.
Therefore $\left[S_{2,2}(-x) - \Li2^2(-x)/2 - c_1\right]/$ $(1-x)$ can be 
represented by a polynomial directly,
where
%--------------------------------------------------------------------------
\begin{equation}
c_1 = 
%%STARTFORM
%%L %% c1const = 
- {7 \over 8} \* \z2^2 + 2 \* \Li4(1/2) 
+ {7 \over 4} \* \z3 \* \lnt 
- {1 \over 2} \* \z2 \* (\lnt)^2 + {1 \over 12} \* (\lnt)^4
%%;
%%STOPFORM
\, .
\end{equation}
%--------------------------------------------------------------------------
We will do this for the functions
$\left[S_{2,2}(-x) - c_1  - \z2^2/8\right]/(1-x)$ and
$\left[\Li2^2(-x) - \z2^2/4\right]/(1-x)$ separately.

The following representation for $\Mvec[(S_{2,2}(-x) - c_1 - \z2^2/8)/(1-x)]$ is obtained:
%------------------------------------------------------------------------
\begin{equation}
\Mvec\left[\frac{S_{2,2}(-x) - c_1 - \z2^2/8}{1-x}\right](N) = \sum_{k=0}^{23} \frac{b^{(1)}_k}{N+k+1}~,
\end{equation}
%------------------------------------------------------------------------
and 
%-----------------------------------------------------------------------
\small
\begin{eqnarray}
%NEW
\begin{array}{lcrlcr}
%%STARTMAPLE
%%L %%f3appr = 
%%{0
b^{(1)}_{0}   &\:=\:&  -0.8778567156865530\ttD1 &
b^{(1)}_{1}   &\:=\:&  -0.8778567156865530\ttD1 
\nonumber\\
b^{(1)}_{2}   &\:=\:&  0.3721432843134470\ttD1 &
b^{(1)}_{3}   &\:=\:& -0.1834122712421077\ttD1 
\nonumber\\
b^{(1)}_{4}   &\:=\:&  0.1030460620911406\ttD1 &
b^{(1)}_{5}   &\:=\:& -0.6362060457158173\ttD2  
\nonumber\\
b^{(1)}_{6}   &\:=\:&   0.4208927186604172\ttD2 &
b^{(1)}_{7}   &\:=\:&  -0.2933929772309200\ttD2  
\nonumber\\
b^{(1)}_{8}   &\:=\:&   0.2130242091464232\ttD2 &
b^{(1)}_{9}   &\:=\:&  -0.1597936899156854\ttD2
\nonumber\\
b^{(1)}_{10}   &\:=\:& 0.1230894254048324\ttD2 &
b^{(1)}_{11}   &\:=\:&-0.9689535635431825\ttD3  
\nonumber\\
b^{(1)}_{12}   &\:=\:& 0.7755737888131504\ttD3 & 
b^{(1)}_{13}   &\:=\:&-0.6263517402742744\ttD3
\nonumber\\
b^{(1)}_{14}   &\:=\:& 0.5031055861991529\ttD3 &
b^{(1)}_{15}   &\:=\:&-0.3922594818742940\ttD3   
\nonumber\\
b^{(1)}_{16}   &\:=\:& 0.2868111320615437\ttD3 &
b^{(1)}_{17}   &\:=\:&-0.1887234737089442\ttD3    
\nonumber\\
b^{(1)}_{18}   &\:=\:& 0.1068980760727356\ttD3 &
b^{(1)}_{19}   &\:=\:&-0.4971730708906830\ttD4  
\nonumber\\
b^{(1)}_{20}   &\:=\:& 0.1797845625225957\ttD4 &
b^{(1)}_{21}   &\:=\:&-0.4695889540721375\ttD5    
\nonumber\\
b^{(1)}_{22}   &\:=\:& 0.7830613264154134\ttD6 &
b^{(1)}_{23}   &\:=\:&-0.6232207394074941\ttD7   
%%};
%%STOPMAPLE
\end{array}
\nonumber
\end{eqnarray}
\normalsize
The corresponding polynomial in $x$ has an accuracy of $2 \times 10^{-15}$ 
in $[0,1]$.

Similarly, for $\Mvec[(\Li2^2(-x) - \z2^2/4)/(1-x)](N)$ one 
obtains:
%------------------------------------------------------------------------
\begin{equation}
\Mvec\left[\frac{\Li2^2(-x) -
\z2^2/4}{1-x}\right](N) = \sum_{k=0}^{11} \frac{b^{(2)}_k}{N+k+1}~,
\end{equation}
%------------------------------------------------------------------------
with
%-----------------------------------------------------------------------
\small
\begin{eqnarray}
%NEW
\begin{array}{lcrlcr}
%%STARTMAPLE
%%L %%f4appr = 
%%{0
b^{(2)}_{0}   &\:=\:&   -0.6764520210934552\ttD0 &
b^{(2)}_{1}   &\:=\:&   -0.6764520137562308\ttD0
\nonumber\\
b^{(2)}_{2}   &\:=\:&    0.3235476094265664\ttD0 &
b^{(2)}_{3}   &\:=\:&   -0.1764446743143206\ttD0
\nonumber\\
b^{(2)}_{4}   &\:=\:&    0.1081940672246993\ttD0 &
b^{(2)}_{5}   &\:=\:&   -0.7181309059958118\ttD1 
\nonumber\\
b^{(2)}_{6}   &\:=\:&    0.4940999469881481\ttD1 &
b^{(2)}_{7}   &\:=\:&   -0.3290941711692155\ttD1
\nonumber\\
b^{(2)}_{8}   &\:=\:&    0.1916664887064280\ttD1 &
b^{(2)}_{9}   &\:=\:&   -0.8589741767655388\ttD2 
\nonumber\\
b^{(2)}_{10}   &\:=\:&   0.2508898780543465\ttD2 &
b^{(2)}_{11}   &\:=\:&  -0.3476710199486832\ttD3
%%};
%%STOPMAPLE
\end{array}
\nonumber
\end{eqnarray}
\normalsize
The corresponding polynomial in $x$ has an accuracy of $2.5 \times 
10^{-11}$ for $x~\epsilon~[0,1]$.

In Table~1 the values of the maximum relative errors are given for the 
first 40 Mellin moments. Depending on the function accuracies better than
$9 \cdot 10^{-10}$ to $8 \cdot 10^{-13}$ are obtained. We also compared the 
representation of the basic functions through numerical inversion of the
Mellin moments by a complex contour integral. In the range  
$x~\epsilon~[10^{-6}, 0.98]$ maximal relative errors of $2~{\rm to}~8~\cdot~10^{-8}$ 
were obtained, irrespective of the value of the basic functions. In some regions
in $x$ even up to several order of magnitude better results were obtained.

\vspace{1.5cm}
\begin{center}
\renewcommand{\arraystretch}{1.3}
\begin{tabular}[h]{||c||c|c||}
\hline \hline %
\multicolumn{1}{||c||}{ } &
\multicolumn{1}{c|}{Moments, $N \leq 40$} &
\multicolumn{1}{c||}{Inversion} \\
\hline\hline
$[\Li4(x)-\zeta_4]/(1-x)$             & $4.0\ttD12$ & $5.1\ttD8$\\
$[S_{2,2}(x)-\zeta_4/4]/(1-x)$        & $8.2\ttD10$ & $5.4\ttD8$\\
$[S_{2,2}(-x)-c_1]/(1-x)$             & $1.8\ttD10$ & $4.5\ttD8$\\
$[\Li2^2(-x)-\zeta_2^2/4]/(1-x)$      & $3.0\ttD10$ & $4.3\ttD8$\\
$\Li4(x)/(1+x)$                       & $1.2\ttD12$ & $8.0\ttD8$\\
$S_{1,3}(x)/(1+x)$                    & $2.2\ttD12$ & $3.3\ttD8$\\
$S_{2,2}(x)/(1+x)$                    & $1.1\ttD12$ & $1.8\ttD8$\\
$\Li2^2(x)/(1+x)$                     & $2.2\ttD12$ & $3.0\ttD8$\\
$[S_{2,2}(-x)-\Li2^2(-x)/2]/(1+x)$    & $7.2\ttD13$ & $3.2\ttD8$\\
\hline \hline
\end{tabular}
\end{center}
\renewcommand{\arraystretch}{1.0}

\vspace{1mm}
\begin{center}
{\sf Table~1: Relative accuracies of the representations.}
\end{center}

%%%%%%%%%%%%%%%%%%%%%%%%%%%%%%%%%%%%%%%%%%%%%%%%%%%%%%%%%%%%%%%%%%%%%%%%
\section{Conclusions}
%%%%%%%%%%%%%%%%%%%%%%%%%%%%%%%%%%%%%%%%%%%%%%%%%%%%%%%%%%%%%%%%%%%%%%%%
%

\vspace{1mm}\noindent
Single scale quantities in massless QED and QCD such as the Wilson 
coefficients and anomalous dimensions can be represented by nested harmonic sums.
Their representation may be reduced to a small set of basic functions 
in addition to Euler's $\psi^{(k)}(N)$ functions, each of the basic functions 
corresponding to one particular single harmonic sums.
This is possible due to algebraic and structural relations between 
the harmonic sums.
Up to the level of the 3--loop anomalous 
dimensions 13 such basic functions occur. We have derived fast and precise numerical 
representations for the eight new functions contributing at the level of the 3--loop 
anomalous dimensions in QCD. In the region 
$x~\epsilon~[10^{-6}, 0.98]$ which is physically most interesting 
their relative accuracy is better than $10^{-7}$. 
If needed, the procedure outlined in the present paper can readily be generalized to a
higher level of accuracy 
at the expense of more coefficients in the representation. 
Due to use of polynomial forms in the analytic continuations 
very fast representations are obtained well suited for the use in QCD 
evolution codes. The basic functions discussed in the present paper form a 
part of the building blocks for the 3--loop coefficient functions and 
other higher order single scale quantities. 

{\tt FORTRAN} routines are available from the authors upon request.

\vspace{3mm}\noindent
{\bf Acknowledgment.}\\
This paper was supported in part by DFG Sonderforschungsbereich
Transregio 9, Computergest\"utzte Theoretische Physik.
S.M. acknowledges the hospitality of the FNAL Theory Division 
during completion of the work.

%\bibliography{/afs/ifh.de/user/m/moch/theorie/tex/Bib/refs.bib}
%\bibliographystyle{/afs/ifh.de/user/m/moch/theorie/tex/Bib/h-elsevier2}

\end{document}